\documentclass[prb,twocolumn,amsmath,amssymb]{revtex4-2}

\usepackage{graphicx}
\usepackage{texdraw}

\usepackage[colorlinks=true, linkcolor=blue, urlcolor=blue, citecolor=blue]{hyperref}

\usepackage[cp1251]{inputenc}
\usepackage[T1,T2A]{fontenc}
\usepackage[ukrainian,english]{babel}

\begin{document}
	
	\title{Classical Heisenberg and XY models on zigzag ladder lattices
		with nearest-neighbor bilinear-biquadratic exchange: Exact solution for the ground-state problem}
	
	\author{Yu.~I. Dublenych}
	\affiliation{Yukhnovskii Institute for Condensed Matter Physics, National
		Academy of Sciences of Ukraine, 1 Svientsitskii Str., 79011
		Lviv, Ukraine\\
		Max Planck Institute for the Physics of Complex Systems, 
		38 N\"{o}thnitzer Str., 01187 Dresden, Germany}
	
	\date{\today}

	\begin{abstract}{An exact and complete solution of the ground-state problem for the classical Heisenberg and XY models with nearest-neighbor bilinear-biquadratic exchange on two- and three-dimensional lattices composed of isosceles triangles is determined with the use of a cluster method. It is shown how the geometric  frustration due to the presence of triangles as	structural units leads to the emergence of a rich phase diagram with incommensurate spiral orderings and their collinear limits, as well as canted and noncoplanar (conical) structures. Surprisingly, there are two different spiral phases with both continuous and discontinuous phase transitions between them. One of these phases is degenerate on two-dimensional partially anisotropic triangular lattice. This degeneracy is lifted on three-dimensional lattices. Canted phase is highly degenerate and this degeneracy persists on three-dimensional lattices.}
	\end{abstract}
	
	\maketitle
	\section{Introduction}
	
	Frustrated magnetic systems on geometrically constrained lattices, such as the triangular and kagom\'e lattices, Shastry-Sutherland lattice, pyrochlore lattice \cite{10.1063/1.2186278, Diep_2020}, 3D honeycomb zigzag ladder lattice \cite{10.21468/SciPostPhysCore.5.3.047} etc. exhibit rich phase diagrams with exotic phases. The emergence of unusual phases is also due to complex interactions between magnetic moments, in particular to biquadratic exchange coupling.
	 
	Classical Heisenberg models with bilinear-biquadratic exchange interaction have been intensively studied during the last decades \cite{PhysRevLett.105.047203, PhysRevB.85.174420, PhysRevB.88.094404, Kartsev2020}. 
	The reason is that there are many magnets, especially 2D ones, where biquadratic exchange interactions play a key role and cannot be neglected (see for instance \cite{Kartsev2020, PhysRevLett.127.247204}).
	
	Ground states of the classical bilinear-biquadratic Heisenberg model on the isotropic triangular lattice were studied in Refs.~\cite{doi:10.1143/JPSJ.76.073704,PhysRevB.85.174420,PhysRevB.88.094404}. In Ref.~\cite{doi:10.1143/JPSJ.76.073704} four phases were identified in this model: a ferromagnetic phase, a nematic-like phase, a 120$^\circ$ phase (triangular phase with 120 degree angle among the nearest moments), and a phase with noncoplanar orientations of spins (see also figure~1 in Ref.~\cite{PhysRevB.85.174420}).
	
	 We studied noncoplanar ground states of classical Heisenberg spins with bilinear-biquadratic interaction on a fully anisotropic triangular lattice \cite{PhysRevB.96.140401}. However, the investigation of the complete phase diagram for such a model with six interaction parameters is a difficult task. Therefore, in this paper, we present a complete phase diagram for a simpler lattice, namely for a partially anisotropic triangular lattice composed of isosceles triangles (a model with four parameter instead of six). We have already investigated ground states for classical Heisenberg spins on lattices of this type but with bilinear interaction only \cite{PhysRevB.93.054415}. Here, in addition to two bilinear interaction parameters, we consider two biquadratic ones. The problem is reduced to the investigation of the ground states for one triangular plaquette. The solution makes it possible to construct global ground states for many lattices composed of identical isosceles triangles including three-dimensional ones.
	
	\begin{figure}[]
		\begin{center}
			\includegraphics[scale = 1.0]{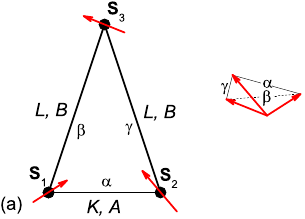}
			
			\vspace{0.15cm}
			
			\includegraphics[scale = 1.5]{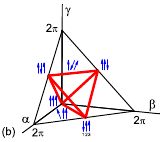}
			\hspace{0.1cm}
			\includegraphics[scale = 0.75]{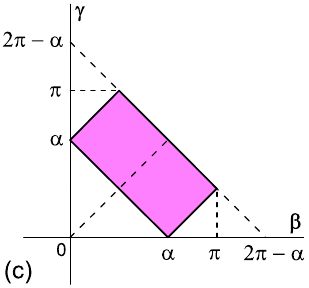}
			\caption{(a) An elementary triangular plaquette for zigzag ladder lattices with spins $\textbf{S}_1$,  $\textbf{S}_2$, and $\textbf{S}_3$ (unit 3-vectors) at its vertices. There are linear ($K$ and $L$) and biquadratic ($A$ and $B$) pairwise interactions (per one plaquette) between the spins. (b) Tetrahedron of values for angles $\alpha$, $\beta$, and $\gamma$ between the spins. Vertices of the tetrahedron correspond to three collinear spins; edges and faces to three coplanar ones; for edges two of three spins are collinear. (c) Cross section of the tetrahedron at a fixed value of $\alpha$.}
			\label{fig1}
		\end{center}
		
	\end{figure}

	\section{Ground-state structures and diagrams for one triangle}
	Let us consider one isosceles triangular plaquette with classical spins (unit 3-vectors) at its vertices [see Fig.\ref{fig1}(a)]. The spins interact via belinear ($K$ and $L$) and biquadratic ($A$ and $B$) pairwise exchange interactions. Interaction parameters are given per one plaquette. If a bond is shared by $n$ plaquettes, then the parameters should be multiplied by $n$ to obtain the interactions between corresponding spins. The Hamiltonian of a zigzag ladder lattice composed of such plaquettes can be presented as a sum over all the plaquettes,
	
	\begin{eqnarray}
		&&H = \sum_{\vartriangle_i} \left\{K\textbf{S}_{i1} \cdot \textbf{S}_{i2}
		+ L(\textbf{S}_{i1} \cdot \textbf{S}_{i3} + \textbf{S}_{i2} \cdot \textbf{S}_{i3})\right.\nonumber\\
		&&\left. -A(\textbf{S}_{i1} \cdot \textbf{S}_{i2})^ 2 - B\left[(\textbf{S}_{i1}
		\cdot \textbf{S}_{i3})^2 + (\textbf{S}_{i2} \cdot \textbf{S}_{i3})^2\right]\right\},
		\label{eq1}
	\end{eqnarray}
	where $\textbf{S}_{i1}$, $\textbf{S}_{i2}$, and $\textbf{S}_{i3}$ are the spins
	at the three vertexes of $i$-th triangular plaquette.
		
	Let $\alpha$, $\beta$, and $\gamma$ are angles between the spins ($0
	\leqslant \alpha,~\beta,~\gamma \leqslant \pi $). These angles
	should satisfy the following inequalities:
	\begin{eqnarray}
		&&~~\alpha + \beta + \gamma \leqslant 2\pi,\nonumber\\
		&&-\alpha + \beta + \gamma \geqslant 0,\nonumber\\
		&&~~\alpha - \beta + \gamma \geqslant 0,\nonumber\\
		&&~~\alpha + \beta - \gamma \geqslant 0.
		\label{eq2}
	\end{eqnarray}
	
	The solution of this set of inequalities is the tetrahedron shown
	in Fig.~\ref{fig1}(b). If at least one of the inequalities becomes an
	equality then the vectors are coplanar. This corresponds to a point
	on the surface of the tetrahedron. The coupling energy between the
	spins of the plaquette shown in Fig.~1(a) reads
	\begin{eqnarray}
		&&E = K\cos \alpha + L(\cos\beta + \cos\gamma)\nonumber\\ 
		&&~~~- A\cos^2\alpha - B(\cos^2\beta + \cos^2\gamma).
		\label{eq3}
	\end{eqnarray}

To obtain the ground-state phase diagram for the $XY$ model, one should minimize the energy (\ref{eq3}) on the tetrahedron surface, i.e. at all vertices, edges, and faces. 
There are 

four vertices $(\alpha, \beta, \gamma)$:
\begin{equation}
	(0, 0, 0); (0, \pi, \pi); (\pi, 0, \pi); (\pi, \pi, 0);
	\label{eq4}
\end{equation}

\begin{table*}[]
	\caption{Ground-state configurations for the classical bilinear-biquadratic Heisenberg model on one isosceles triangle and their energies. The first, second and third groups of rows correspond, respectively, to the vertices, edges, and faces of the tetrahedron shown in Fig.~\ref{fig1} [see also Eqs.~(\ref{eq4}--\ref{eq6})]. The last row corresponds to the interior points of the tetrahedron.}%
	\begin{center}
		\begin{tabular}{clcl}
			\hline \hline
			Angles $\alpha$, $\beta$, $\gamma$&\multicolumn{1}{c}{Energy~~~}&Conditions&Structure\\
			\hline&\\[-2mm]
			$\alpha = \beta = \gamma = 0$&~~~$E = K + 2L - A - 2B$&&\includegraphics[scale = 0.8]{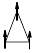}~~F\\[2mm]
			$\alpha = 0$, $\beta = \pi$, $\gamma = \pi$&~~~$E = K - 2L - A - 2B$&&\includegraphics[scale = 0.8]{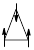}~~AF$_1$\\[2mm]
			$\alpha = \pi$, $\beta = 0$, $\gamma = \pi$&~~~$E = -K - A - 2B$&&\includegraphics[scale = 0.8]{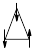}~~AF$_{21}$\\
			$\alpha = \pi$, $\beta = \pi$, $\gamma = 0$&~~~$E = -K - A - 2B$&&\includegraphics[scale = 0.8]{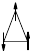}~~AF$_{22}$\\[2mm]
			\hline\\
			$\alpha = 0$, $\cos\beta = \frac{L}{2B}$, $\gamma = \beta$&~~~$E = K - A + \frac{L^2}{2B}$&~~~$|L| <  2|B|$&\includegraphics[scale = 0.8]{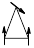}~~CF\\[2mm]
			$\cos\alpha = \frac{K + L}{2(A + B)}$, $\beta = 0$, $\gamma = \alpha$&~~~$E = \frac{(K + L)^2}{4(A + B)} + L - B$&~~~$|K + L| < 2|A + B|$&\includegraphics[scale = 0.8]{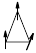}~~CF$_{11}$\\
			~~~~~~~~~~~~~~~~~~or $\beta = \alpha$, $\gamma = 0$&&&\includegraphics[scale = 0.8]{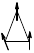}~~CF$_{12}$\\[2mm]
			$\cos\alpha = \frac{K - L}{2(A + B)}$, $\beta = \pi$, $\gamma = \pi- \alpha$&~~~$E = \frac{(K - L)^2}{4(A + B)} - L - B$&~~~$|K - L| < 2|A + B|$&\includegraphics[scale = 0.8]{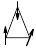}~~CF$_{21}$\\
			~~~~~~~~~~~~~~~~~~or $\beta = \pi- \alpha$, $\gamma = \pi$&&&\includegraphics[scale = 0.8]{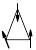}~~CF$_{22}$\\[2mm]		
			$\alpha = \pi$, $\beta = \frac{\pi}{2}$, $\gamma = \frac{\pi}{2}$&~~~$E = -K - A$&&	\includegraphics[scale = 0.8]{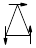}~~CF$_3$\\[2mm]
			\hline\\			
			$8A\cos^3\frac{\alpha}{2} + 2(B-K-2A)\cos\frac{\alpha}{2} - L = 0$,&~~~$E = 4A\cos^4\frac{\alpha}{2} +L\cos\frac{\alpha}{2}- K - A$&~~~$0 < \cos\frac{\alpha}{2} < 1$&\includegraphics[scale = 0.8]{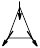}~~Sp$_{11}$\\
			$\beta = \gamma = \frac{\alpha}{2}$&&&\\
			$8A\cos^3\frac{\alpha}{2} + 2(B-K-2A)\cos\frac{\alpha}{2} + L = 0$,&~~~$E = 4A\cos^4\frac{\alpha}{2} - L\cos\frac{\alpha}{2}- K - A$&~~~$0 < \cos\frac{\alpha}{2} < 1$&\includegraphics[scale = 0.8]{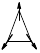}~~Sp$_{12}$\\
			$\beta = \gamma = \pi - \frac{\alpha}{2}$&&&\\[2mm]
			
			$8AB \cos^3\alpha - 4B(K + B)\cos^2\alpha + L^2 = 0$,&~~~$E = A\cos^2\alpha + \frac{L^2}{2B\cos\alpha}+\frac{L^2}{4B} - B$&$0 < \alpha \leqslant \arccos \frac{L^2 + \sqrt{L^4 + 32B^2L^2}}{16B^2}$&\includegraphics[scale = 0.8]{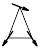}~~Sp$_{21}$\\
			$\beta$ is determined from Eq. (\ref{eq16}),&&or&\includegraphics[scale = 0.8]{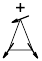}~~Sp$_{22}$\\
			$\gamma$ is determined from Eqs. (\ref{eq6})&&$\arccos \frac{L^2 - \sqrt{L^4 + 32B^2L^2}}{16B^2} \leqslant \alpha < \pi$&\\[3mm]
			\hline\\				
			$\cos\alpha = \frac{K}{2A}$, $\cos\beta = \frac{L}{2B}$, $\gamma = \beta$&~~~$E = \frac{K^2}{4A}+\frac{L^2}{2B}$&~~~$|K| < 2|A|$, $|L| < 2|B|$&Noncoplanar\\[3mm]
			
			\hline \hline
		\end{tabular}
	\end{center}
	\label{table1}
\end{table*}

six edges:
\begin{eqnarray}
	&&\alpha = 0,~ \gamma = \beta;~~ \beta = 0,~ \gamma = \alpha;~~ \gamma = 0,~ \beta = \alpha;\nonumber\\
	&&~~~~~~~~~~~~~~~~~~\alpha + \beta = \pi,~ \gamma = \pi;\nonumber\\
	&&~~~~~~~~~~~~~~~~~~\alpha + \gamma = \pi,~ \beta = \pi;\nonumber\\
	&&~~~~~~~~~~~~~~~~~~\beta + \gamma = \pi,~ \alpha = \pi;
	\label{eq5}
\end{eqnarray}

and four faces:
\begin{eqnarray}
	&&-\alpha + \beta + \gamma = 0,\nonumber\\
	&&~~\alpha - \beta + \gamma = 0,\nonumber\\
	&&~~\alpha + \beta - \gamma = 0,\nonumber\\
	&&~~\alpha + \beta + \gamma = 2\pi.
	\label{eq6}
\end{eqnarray}

	\begin{figure*}[]
	\begin{center}
		\includegraphics[scale = 0.75]{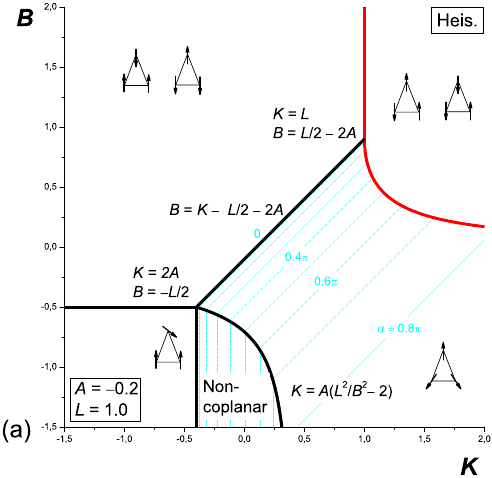}
		\hspace{1cm}
		\includegraphics[scale = 0.75]{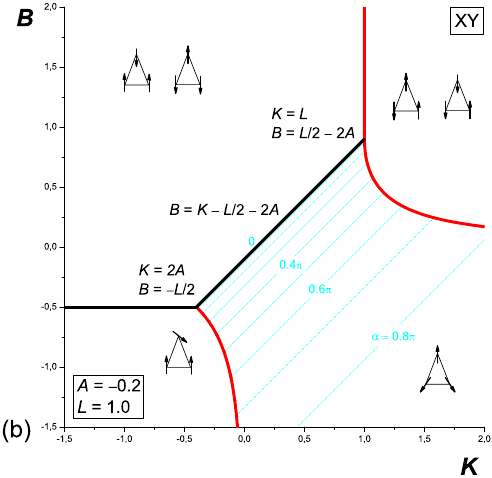}
		\vspace{0.5cm}
		\includegraphics[scale = 0.75]{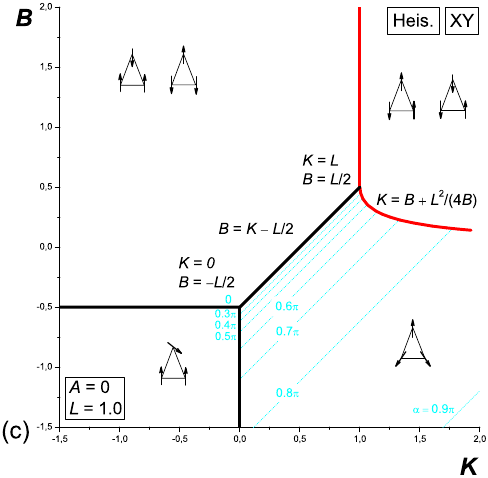}
		\hspace{1cm}
		\includegraphics[scale = 0.75]{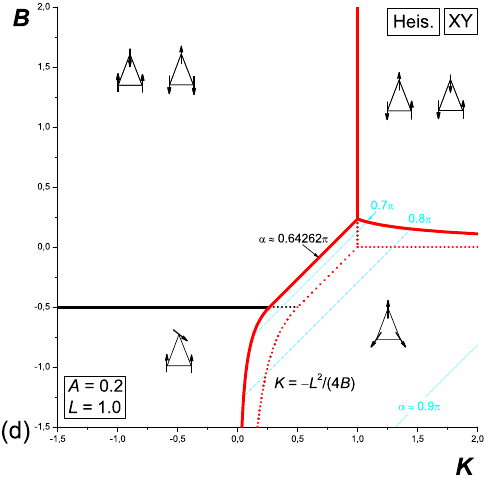}
		\caption{Ground-state phase diagrams for the bilinear-biquadratic Heisenberg and XY models on one isosceles  triangle. (a) $A = -0.2|L|$ (Heisenberg model), (b) $A = -0.2|L|$ (XY model), (c) $A = 0$ (both models), (d) $A = 0.2|L|$ (both models). Black and red lines correspond to continuous and discontinuous  transitions, respectively. Dotted lines correspond to $A \rightarrow +\infty$. The diagram of type (a) for the Heisenberg model occurs if $-\frac{4}{9}\sqrt6 L \approx -1.08866 L < A < 0$ and the diagram of type (b) for XY model occurs if $-\frac12 < A < 0$.}
		\label{fig2}
	\end{center}
\end{figure*}
	
	If the energy attains its minimum in an intrinsic point of the
	tetrahedron [Fig.~1(b)], then this minimum is determined from the
	following conditions:
	\begin{eqnarray}
		&&\frac{\partial E}{\partial \alpha} = (-K + 2A\cos\alpha)\sin\alpha = 0,\nonumber\\
		&&\frac{\partial E}{\partial \beta} = (-L + 2B\cos\beta)\sin\beta = 0,\nonumber\\
		&&\frac{\partial E}{\partial \gamma} = (-L + 2B\cos\gamma)\sin\gamma = 0.
		\label{eq7}
	\end{eqnarray}

It follows from these equations
	\begin{equation}
		\cos\alpha = \frac{K}{2A}, \cos\beta = \cos\gamma = \frac{L}{2B}.
		\label{eq8}
	\end{equation}

	\begin{figure*}[]
		\begin{center}
			\includegraphics[scale = 0.9]{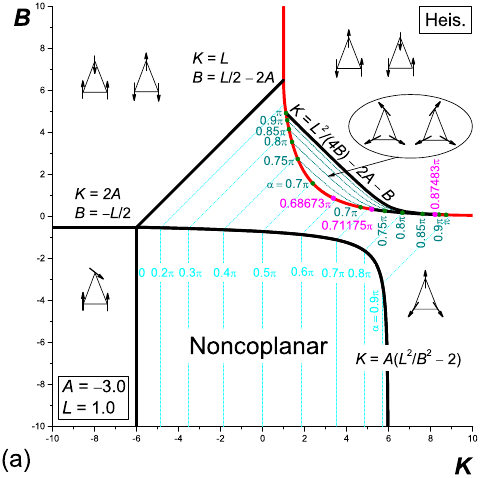}
			\hspace{1cm}
			\includegraphics[scale = 0.9]{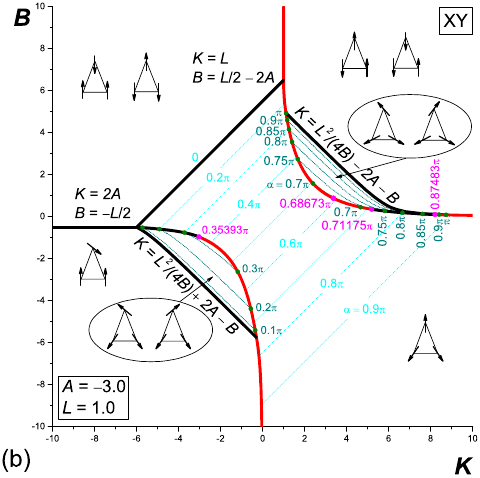}
			\vspace{0.5cm}
			\includegraphics[scale = 0.9]{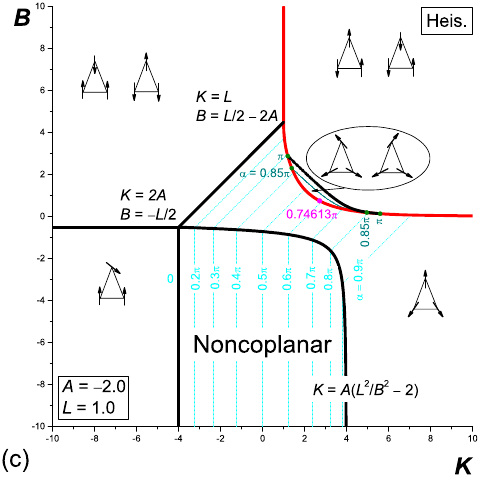}
			\hspace{1cm}
			\includegraphics[scale = 0.9]{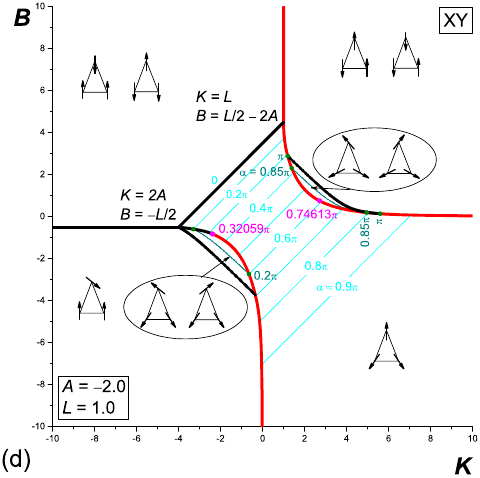}
	\caption{Ground-state phase diagram for the bilinear-biquadratic (a), (c) Heisenberg and (b), (d) XY models on one isosceles triangle at $A = -3.0$ and $A = -2.0$ ($L = 1.0$). Diagrams of these types occur when $A < -\frac{4}{9}\sqrt6 L \approx -1.08866 L~~(L > 0)$. Black and red lines correspond to continuous and discontinuous  transitions, respectively. The tricritical points and the points of minimum $\alpha$ value for Sp$_2$ structure are indicated with magenta color. At $A = -3.0$ and $L = 1.0$, $\alpha \approx 0.68673486445\pi$ [see Figs. (a) and (b)].}
	\label{fig3}
	\end{center}
	\end{figure*}

	\begin{figure}[]
		\begin{center}
			\includegraphics[scale = 0.75]{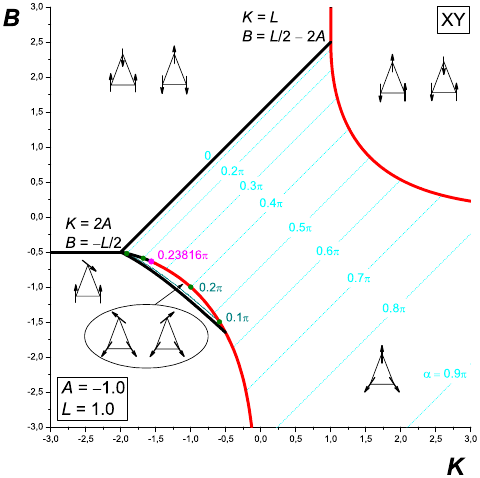}		
			\caption{Ground-state phase diagram for the XY model on one isosceles triangle at $A = -1.0|L|$. Diagrams of these types arise when $-\frac{4}{9}\sqrt6 |L| < A < -\frac12|L|$.}
			\label{fig4}
		\end{center}
	\end{figure}

	Let us find the minimum of the energy (\ref{eq3}) on the face $\gamma = \alpha - \beta$ of the tetrahedron [see Fig.~\ref{fig1}(b)].
			
	\begin{eqnarray}			
		&&\frac{\partial E}{\partial \alpha} = \sin\alpha \left[2A\cos\alpha - K\right]  \nonumber\\
		&& + \sin(\alpha - \beta)\left[2B\cos(\alpha - \beta) - L\right] = 0, \nonumber\\
		&&\frac{\partial E}{\partial \beta} = L\left[\sin(\alpha - \beta) - \sin\beta\right] \nonumber\\
		&& - B\left[\sin2(\alpha-\beta) - \sin2\beta\right] = 0.
		\label{eq9}
	\end{eqnarray}

We have from the last equation

	\begin{equation}
		L\sin\left(\frac{\alpha}{2} - \beta\right)\cos\frac{\alpha}{2} -
		B\sin(\alpha - 2\beta)\cos\alpha = 0
		\label{eq10}
	\end{equation}
	or
	\begin{equation}
		\sin\left(\frac{\alpha}{2} - \beta\right)\left[L\cos\frac{\alpha}{2} -
		2B\cos\left(\frac{\alpha}{2} - \beta\right)\cos\alpha\right] = 0.
		\label{eq11}
	\end{equation}
	
	This equation yelds
		\begin{equation}
		\beta = \frac{\alpha}{2}~\text{or}~L\cos\frac{\alpha}{2} -
		2B\cos\left(\frac{\alpha}{2} - \beta\right)\cos\alpha = 0.
		\label{eq12}
	\end{equation}

In the same way, we have for the face $\alpha + \beta + \gamma = 2\pi$
	
	\begin{equation}
		\beta = \pi - \frac{\alpha}{2}~\text{or}~L\cos\frac{\alpha}{2} -
		2B\cos\left(\frac{\alpha}{2} + \beta\right)\cos\alpha = 0.
		\label{eq13}
	\end{equation}

For faces $\alpha - \beta + \gamma = 0$ and $\alpha + \beta - \gamma = 0$, equations are the same as for faces $-\alpha + \beta + \gamma = 0$ [Eq. (\ref{eq12})] and $\alpha + \beta + \gamma = 2\pi$ [Eq. (\ref{eq13})], respectively.

Let us solve the equation
\begin{equation}
	L\cos\frac{\alpha}{2} -
	2B\cos\left(\frac{\alpha}{2} \pm \beta\right)\cos\alpha = 0.
	\label{eq14}
\end{equation}

After simple transformations we obtain a quadratic equation with respect to $\cos\beta$
\begin{eqnarray}
	&&4B^2\sin^2\frac{\alpha}{2}(1-\cos^2\beta)\cos^2\alpha \nonumber\\
	&&= \left(L\cos\frac{\alpha}{2} - 2B\cos\frac{\alpha}{2}\cos\alpha\cos\beta\right)^2.
	\label{eq15}
\end{eqnarray}

It has the solution
\begin{eqnarray}
	&&\cos\beta = \frac{L(1+\cos\alpha)}{4B\cos\alpha}\nonumber\\
	&&\pm\frac{\sqrt{(\cos\alpha -1)[L^2(1+\cos\alpha)-8B^2\cos^2\alpha]}}{4B\cos\alpha}
	\label{eq16}
\end{eqnarray}
at the condition
\begin{equation}
	8B^2\cos^2\alpha - L^2\cos\alpha - L^2 \geqslant 0,
	\label{eq17}
\end{equation}
that is equivalent to the conditions
\begin{eqnarray}
	&&0 < \alpha \leqslant \arccos \frac{L^2 + \sqrt{L^4 + 32B^2L^2}}{16B^2}, ~~ |L| \leqslant 2|B|; \nonumber\\
	&&\arccos \frac{L^2 - \sqrt{L^4 + 32B^2L^2}}{16B^2} \leqslant \alpha < \pi.
\label{eq18}
\end{eqnarray}

For all the faces, the expression for the energy can be written in the following form: 
	\begin{eqnarray}
		&&E = K\cos\alpha - A\cos^2\alpha + L(\cos\beta +\cos(\alpha \pm \beta)) \nonumber\\
		&&-B(\cos^2\beta + \cos^2(\alpha \pm \beta)) \nonumber\\
		&&= K\cos\alpha - A\cos^2\alpha +2L\cos\frac{\alpha}{2}\cos\left(\frac{\alpha}{2} \pm \beta\right) \nonumber\\
		&&-B\left[1 + \cos\alpha\left(2\cos^2\left(\frac{\alpha}{2} \pm \beta\right) - 1\right)\right].
		\label{eq19}
	\end{eqnarray}

Equation \ref{eq12} yields
	\begin{equation}
		\cos\left(\frac{\alpha}{2} \pm \beta\right) = \frac{L\cos\frac{\alpha}{2}}{2B\cos\alpha}.
		\label{eq20}
	\end{equation}

Using this equation, the expression (\ref{eq19}) can be written as 
	\begin{eqnarray}
		&&E = - A\cos^2\alpha + K\cos\alpha + B(\cos\alpha - 1) \nonumber \\
		&&~~~~~+ \frac{L^2(\cos\alpha + 1)}{4B\cos\alpha}.
		\label{eq21}
	\end{eqnarray}

Finding the derivative of this function with respect to $\alpha$,
	\begin{equation}
		\frac{\partial E}{\partial \alpha} = \frac{\sin\alpha\left(8AB\cos^3\alpha - 4B(B+K)\cos^2\alpha + L^2\right)}{4B\cos^2\alpha},
		\label{eq22}
	\end{equation}
we obtain the equation
	\begin{equation}
		8AB\cos^3\alpha - 4B(B+K)\cos^2\alpha + L^2= 0.
		\label{eq23}
	\end{equation}

Taking into account this equation, we can write the expression for the energy in the following form:
	\begin{equation}
		E = A\cos^2\alpha + \frac{L^2}{2B\cos\alpha} + \frac{L^2}{4B} - B.
		\label{eq24}
	\end{equation}
Equations (\ref{eq23}) and (\ref{eq24}) are the same for all the faces.

In a similar way we obtain the equations for $\alpha$ and the expressions for the energy for $\beta = \frac{\alpha}{2}$ and $\beta = \pi - \frac{\alpha}{2}$ [see Eqs. (\ref{eq12}) and (\ref{eq13})].

\begin{figure*}[]
	\begin{center}
		\includegraphics[scale = 0.65]{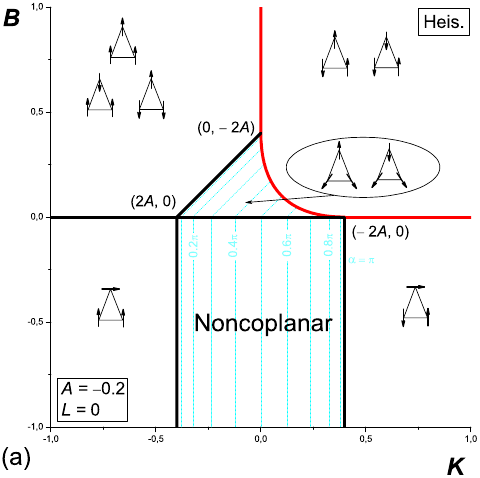}
		\hspace{0.5cm}
		\includegraphics[scale = 0.65]{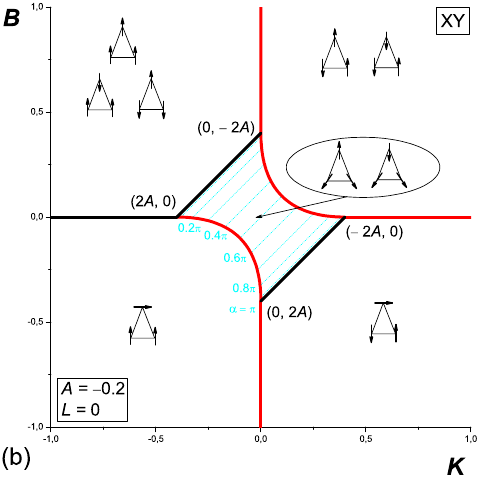}	
		\hspace{0.5cm}
		\includegraphics[scale = 0.65]{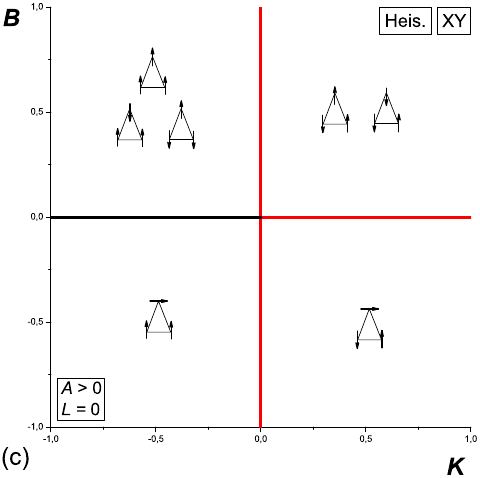}			
		\caption{Ground-state phase diagram for the bilinear-biquadratic (a), (c) Heisenberg and (b), (c) XY models on one isosceles triangle at $L = 0$.}
		\label{fig5}
	\end{center}
\end{figure*}

All these results are presented in Table~\ref{table1}. The first, second and third groups of rows correspond, respectively, to the vertices, edges, and faces of the tetrahedron show in Fig.~\ref{fig1} [see also Eqs.(\ref{eq4}--\ref{eq6})]. The last row corresponds to the interior points of the tetrahedron and has to be considered only for the Heisenberg model. In the first column of the Table~\ref{table1}, three angles, $\alpha$, $\beta$, and $\gamma$, between three vectors are given. In the third group, $\cos\alpha$ (or $\cos\frac{\alpha}{2}$) is determined as a solution of a cubic equation. In the second column, the energy (per one plaquette) of the spin structure is given as a function of the interaction parameters and the angle $\alpha$. In the third column, necessary conditions for existence of the ground-state structure are indicated and, in the last column, the triangular plaquette spin structures are shown. Denotations F, AF, CF, and Sp mean ``Ferromagnetic,'' ``Antiferromagnetic,'' ``Canted Ferromagnetic,'' and ``Spiral,'' respectively. 

It should be noted that all three spins can be rotated by the same angle in the 3D-space, since only the angles between pairs of spin matter. One can also swap the spins at the base of the isosceles triangle. The angles also do not change if all three spins are replaced by opposite ones. In the structures with two collinear spins (CF structures), one can rotate the third spin by an arbitrary angle with respect to the axis of the collinear ones. 

\begin{figure}[]
	\begin{center}
		\includegraphics[scale = 0.9]{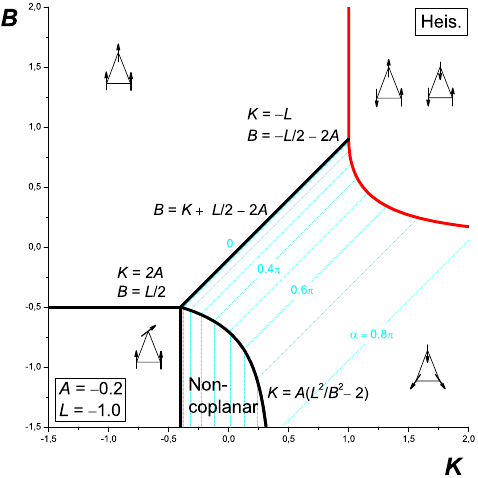}		
		\caption{Ground-state phase diagram for the bilinear-biquadratic Heisenberg model on one isosceles triangle at $A = -0.2$, $L = -1.0$}
		\label{fig6}
	\end{center}
\end{figure}

Using Table~\ref{table1} one can construct the ground-state phase diagrams for the bilinear-biquadratic Heisenberg and $XY$ models. We have constructed these diagrams in $(K, B)$-plane. They are shown in Figs.~\ref{fig2}--\ref{fig6}. It is sufficient to investigate the ground-state phase diagrams for $L = 1$ and $L = 0$ only. For $L = -1$ the diagrams are similar to those for $L = 1$. One just need to replace structure AF$_1$ with structure F, structures CF$_{21}$ and CF$_{22}$ with structures CF$_{11}$ and CF$_{12}$, and structures Sp$_{12}$ with structures Sp$_{11}$ [see Table~\ref{table1} and Figs.~\ref{fig2}(a) and \ref{fig6}].

The ground-state phase diagrams in the $(K, B)$-plane for $A > -\frac{4}{9}\sqrt6 L \approx -1.08866 L~(L > 0)$ are shown in Fig.~\ref{fig2}. In this case, there are four ground-state structures: collinear, AF$_1$, AF$_2$ (AF$_{21}$ and AF$_{22}$), coplanar, CF, Sp$_{12}$, and, for the Heisenberg model at $A < 0$, a noncollinear. At the boundary of two phases the angles $\alpha$, $\beta$, and $\gamma$ can change continuously or discontinuously. In Figs.~\ref{fig2}--\ref{fig6}, lines of discontinuous transitions are shown with red color. Transitions between some phases can be both continuous and discontinuous, depending on values of parameters. In this case, there are tricritical points, where continuous and discontinuous transitions meet.

At $A < -\frac{4}{9}\sqrt6 L~(L > 0)$, in addition to the phases listed above, there is the phase Sp$_2$ (Sp$_{21}$ and Sp$_{22}$). Between phases Sp$_2$ and AF$_2$ and between phases Sp$_2$ and CF (for the XY model) the transition is continuous.

In the Heisenberg model, transitions between phases Sp$_2$ and Sp$_{12}$ are both continuous and discontinuous. Continuous transitions exist if $A < -\frac{25\sqrt{10}}{32} L \approx -2.4705 L$ ($L > 0$) (see Appendix). In this case, at the boundary of the phases, there are two tricritical points and the point of minimal value of the angle $\alpha$ for the phase Sp$_2$ at fixed $A$ and $L$ [Fig.~\ref{fig3}(a)].

Phases CF$_1$ and CF$_2$ cannot have minimum energy in any region of the model parameter space. The phase CF$_3$ can have minimum energy at $L = 0$ only (Fig.~\ref{fig5}).

\begin{figure*}[]
	\begin{center}
		\includegraphics[scale = 1.2]{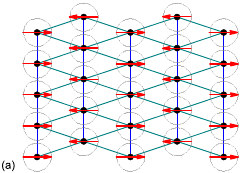}
		\hspace{0.5cm}
		\includegraphics[scale = 1.2]{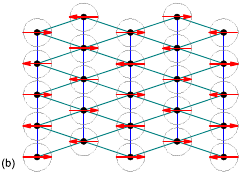}
		\hspace{0.5cm}
		\includegraphics[scale = 1.2]{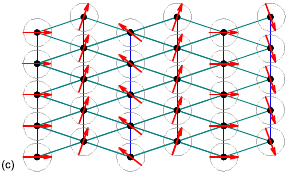}
		\vspace{0.5cm}
		\includegraphics[scale = 1.2]{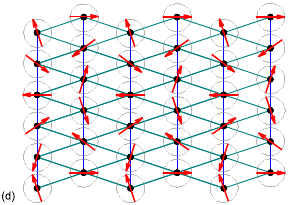}
		\hspace{0.5cm}
		\includegraphics[scale = 1.2]{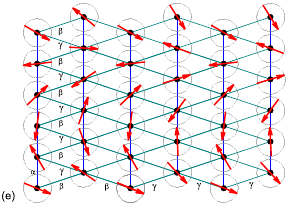}		
		\caption{Ground-state structures for a partially anisotropic triangular lattice with a nearest-neighbor bilinear-biquadratic exchange. The interaction parameters are equal to $2K$, $2L$, $2A$, and $2B$ (see Fig.~\ref{fig1}). (a), (b) Antiferromagnetic AF$_1$ and AF$_2$ structures. (c) Canted CF structures ($\beta = \gamma = 70^\circ$). All the spins are coplanar, however for the Heisenberg model they can be noncoplanar in general case. (d) Periodic spiral Sp$_{12}$ structure with $\alpha = 0.8 \pi$, $\beta = \gamma = \pi - \frac{\alpha}{2} = 0.6\pi$. The magnetization is equal to zero. All Sp$_{12}$ structures are periodic in the direction perpendicular to the ladder legs with period of two ladders, but not necessarily periodic in the direction of the legs. (e) Incommensurate spiral Sp$_{2}$ structure for  $K = 3.5$, $L = 1.0$, $A = -3.0$, $B = 1.5$. $\alpha \approx 142.06^\circ$, $\beta \approx 168.93^\circ$, $\gamma \approx 26.87^\circ$.} 
		\label{fig7}
	\end{center}
\end{figure*}

\section{Global ground-state structures on zigzag ladder latticies}

\begin{figure}[]
	\begin{center}
		\includegraphics[scale = 1.2]{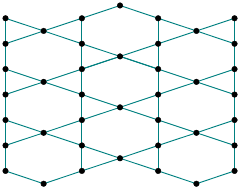}			
		\caption{Distorted kagome lattice is composed with corner sharing isosceles triangles. This is not a zigzag ladder lattice. A part of ground-state spin configurations for this lattice can be obtained from anisotropic triangular lattice configurations by removing some of the lattice sites. The interaction parameters are equal to $K$, $L$, $A$, and $B$.}
		\label{fig8}
	\end{center}
\end{figure}

\begin{figure}[]
	\begin{center}
		\includegraphics[scale = 1.2]{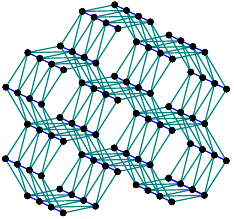}			
		\caption{3D honeycomb zigzag ladder lattice. The interaction parameters are equal to $3K$, $2L$, $3A$, and $2B$.}
		\label{fig9}
	\end{center}
\end{figure}

\begin{figure}[]
	\begin{center}
		\includegraphics[scale = 0.75]{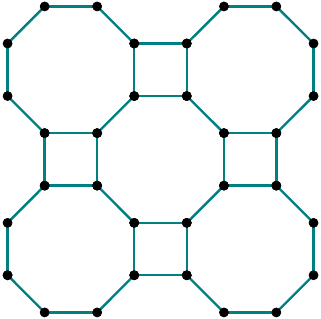}			
		\caption{A 3D zigzag ladder lattice can be also constructed on the base of 4.8$^2$ Euclidean tiling. The interaction parameters are equal to $3K$, $2L$, $3A$, and $2B$.}
		\label{fig10}
	\end{center}
\end{figure}

Using the ground-state structures and diagrams for one triangle, we can construct global ground-state structures for partially anisotropic triangular lattice. They are shown in Fig.~\ref{fig7}. Whatever the directions of the spins, the angles between spins should be the same for all the triangular plaquettes. Since each pair of spins belongs to two neighboring plaquettes on the lattice, the interaction parameters are equal to $2K$, $2L$, $2A$, and $2B$. For a distorted kagome lattice (Fig.~\ref{fig8}) the interaction parameters remain the same as for one triangle. We can also construct ground-state structures for three-dimensional zigzag ladder lattices on the base of planar bipartite lattices (see Figs.~\ref{fig9} and \ref{fig10} as examples).

The CF spin structure on one triangle gives rise to a highly degenerate global phase on zigzag ladder lattice (Fig.~\ref{fig7}(c)). All the spins on the same leg should have the same direction but only the angle between the spins on adjacent legs matters. Thus, giving a spin configuration on one leg of a zigzag ladder, there are two ways to construct the spin configuration for the other leg in the XY model and infinite number of ways in the Heisenberg model. Noncollinear global CF structures for the Heisenberg model are possible and this leads to a high degeneracy. This degeneracy is not lifted on 3D latitces.

\begin{figure}[]
	\begin{center}
		\includegraphics[scale = 1.5]{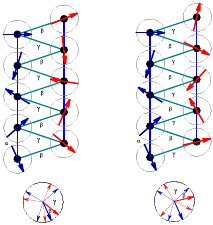}
		\caption{Two ways to construct an adjacent spin chain on the zigzag ladder. The interaction parameters  are equal to $K$, $2L$, $A$, $2B$.}
		\label{fig11}
	\end{center}
\end{figure}

Let us consider the coplanar Sp$_2$ spin structure. Giving a spin configuration on one leg of a zigzag ladder, there are in general two ways to construct the spin configuration on the other leg. This is shown in Fig.~\ref{fig11}. Thus, the global Sp$_2$ spin structure is nontrivially degenerate. This degeneracy is lifted for 3D zigzag ladder lattices, for instance, the honeycomb zigzag ladder lattice (Fig.~\ref{fig9}) or a zigzag ladder lattice based on the 4.8$^2$ (Fig.~\ref{fig10}), 4$^4$ (square lattice), and 4.6.12 Euclidean tiling.

A distorted kagome lattice  (Fig.~\ref{fig8}) is composed with corner sharing isosceles triangles although this is not a zigzag ladder lattice. A part of ground-state spin configurations for this lattice can be obtained from anisotropic triangular lattice configurations by removing every forth site. But some ground states of the kagome lattice are more degenerate, for instance CF structure. The interaction parameters are equal to $K$, $L$, $A$, and $B$.

All Sp$_{12}$ structures are periodic in the direction perpendicular to the ladder legs with a period of two ladders, but not necessarily periodic in the direction of the legs. Thus, these structures generate ground-state structures for 3D zigzag ladder lattices as well. 

The Sp$_2$ phase is degenerate because the sequence of $\beta$ and $\gamma$ that determines a ground-state structure [see the lower part of Fig.~\ref{fig7}(e)] can be arbitrary. And only one sequence, $...\beta$, $\beta$, $\beta$,... (or ...$\gamma$, $\gamma$, $\gamma$, ...), gives a periodic structure with a period of two ladders and therefore generates a ground-state structure for 3D zigzag ladder lattices.

We do not consider here the noncoplanar ground-state phase. This was done for the isotropic triangular lattice in Ref.~\cite{PhysRevB.85.174420} and for the fully anisotropic triangular lattice in Ref.~\cite{PhysRevB.96.140401}.

\section{Conclusions}
We have investigated an exact and complete ground-state phase diagram for the classical Heisenberg and XY models on various two- and three-dimensional zigzag ladder lattices with nearest-neighbor bilinear-biquadratic exchange. The problem is reduced to constructing ground-state phase diagram for one triangular plaquette. Nevertheless, two different biquadratic-exchange parameters together with two bilinear-exchange ones give rise to a complex ground-state phase diagram with collinear, coplanar, and noncoplanar spin structures. There are continuous and discontinuous transitions between phases. The most interesting result is the existence of two different spiral phases (Sp$_1$ and Sp$_2$) between which there are both continuous and discontinuous transitions, and therefore, tricritical points at the boundary. The spiral phase Sp$_2$ on the anisotropic triangular lattice is non-trivially degenerate; the degeneracy is lifted on 3D lattices. The canted phase CF is highly degenerate on the anisotropic triangular lattice in the Heisenberg model and this degeneracy is not lifted on 3D lattices.

\section{Acknowledgment}
The author is gratefull for the hospitality of the Max Planck Institute for the Physics of Complex Systems where the work has been completed. 

	\section*{Appendix}

We show here how to construct the ground-state phase diagrams for the model considered, that is we give equations for boundaries between different phases and for specific points on these boundaries. We also show for some triangular plaquette spin configurations that there are no regions in the parameter space where they have a minimum energy. 

\section*{$\mathbf{A < 0}$, $\mathbf{L > 0}$ case}
  
The equation for \textit{the boundary between} AF$_2$ \textit{and} Sp$_{12}$ \textit{phases} reads
\begin{eqnarray}
	&&B = -2A\cos^4\frac{\alpha}{2} + \frac{L}{2}\cos\frac{\alpha}{2},\nonumber\\
	&&K = -2A\left(\cos^2\frac{\alpha}{2} - 1\right)^2 + \frac{L\left(\cos^2\frac{\alpha}{2} + 1\right)}{2\cos\frac{\alpha}{2}}.
	\label{eq25}
\end{eqnarray}
	
Set of equations for \textit{the point where} Sp$_{12}$, Sp$_2$, \textit{and} CF \textit{phases meet} (XY model) 

	\begin{eqnarray}
		&&16A^2 x^6 - 48A^2 x^4 - 8ALx^3 - 6ALx + L^2 = 0,\nonumber \\
		&&B=\frac{3AL^2x^2}{32A^2x^4 + 4ALx^3 + 4ALx-L^2}, \nonumber \\
		&&K=\frac{L^2}{4B} + 2A - B;~~~~x=\cos\frac{\alpha}{2}.
		\label{eq26}
	\end{eqnarray}

\textit{The boundary between} Sp$_{12}$ \textit{and} Sp$_2$ \textit{phases} is determined from the following set of equations
\begin{eqnarray}
	&&8AB\cos^3\alpha - 4B(K + B)\cos^2\alpha + L^2 = 0, \nonumber \\
	&&8Ax^3 + 2(B - K - 2A)x + L = 0, \\
	&&A\cos^2\alpha + \frac{L^2}{2B\cos\alpha}+\frac{L^2}{4B} - B = 4Ax^4 - Lx - A - K.\nonumber
	\label{eq27}
\end{eqnarray}

The equation for \textit{the line of second order phase transition between} Sp$_{12}$ \textit{and} Sp$_2$ \textit{phases} (XY model) 
	\begin{eqnarray}
		&&B = -\frac{Lx}{2(2x^2-1)},~~x = \cos\frac{\alpha}{2}, \nonumber \\
		&&K = \frac{16Ax^5-16Ax^3+Lx^2+4Ax-L}{2x(2x^2-1)}.
	\label{eq28}
	\end{eqnarray}

Equations for the \textit{tricritical points} read
	\begin{eqnarray}
		&&\cos\alpha = \sqrt[3]{\frac{L^2}{4AB}}, \nonumber \\
		&&B=\frac{L^2}{4A\cos^3\alpha},~~K=3A\cos\alpha - \frac{L^2}{4A\cos^3\alpha}, \nonumber \\
		&&K = 3A\sqrt[3]{\frac{L^2}{4AB}} - B.
		\label{eq29}
	\end{eqnarray}
Values of $\alpha$ for the tricritical points are solutions of the following equation
	\begin{equation}
		2A\left(2\cos^2\frac{\alpha}{2}-1\right)^2\cos\frac{\alpha}{2} + L = 0. 
		\label{eq30}
	\end{equation}
If $\frac{A}{L}< -\frac{25\sqrt{10}}{32} \approx -2.4705$ then this equation has three real positive roots less that 1 and therefore there are three tricritical points in the $(K,B)$-plain for the XY model and two for the Heisenberg model.

The equation for \textit{the boundary between} CF \textit{and} Sp$_{12}$ reads
\begin{eqnarray}
	&&B = \frac{4Ax(x^2-1)^2 - L(x^2+1)}{4x}\nonumber \\
	&&+\frac{(x^2-1)\sqrt{(4Ax(x+1)^2-L)(4Ax(x-1)^2-L)}}{4x}, \nonumber \\
	&&K = \frac{8ABx^4 - 2BLx - L^2}{4B}; ~~x = \cos\frac{\alpha}{2}.
	\label{eq31}
\end{eqnarray}

\textit{The line of first order phase transition between} Sp$_{12}$ \textit{and} Sp$_2$ \textit{phases} (XY model) is determined by the equation 
	\begin{eqnarray}
		&&16A^2x^6 - 16A^2(y^2 + y + 1)x^4  \nonumber \\ 
		&& - 8ALx^3 + 4A^2(-2y^3 - y^2 + 2y + 1)x^2 \nonumber \\
		&& + 2AL(y^2 - 2y - 2)x + L^2 = 0, \nonumber \\
		&&B = \frac{L^2(y + 2)x}{2y[8Ax^5 - 8Ax^3 - 2Lx^2 + 2Ax(1 - y^2) - L]}, \nonumber \\
		&&K=\frac{8Ax^3 - 2(2A - B)x + L}{2x}; ~~0 < x = \cos\frac{\alpha_1}{2}  \leqslant 1,\nonumber \\
		&&y = \cos\alpha_2,~~~-1 < y \leqslant \frac{L^2 - \sqrt{L^4 + 32B^2L^2}}{16B^2} \nonumber \\
		&&~~~~~~~~~~~~~\text{or}~~ \frac{L^2 + \sqrt{L^4 - 32B^2L^2}}{16B^2} \leqslant y < 1.  
		\label{eq32}
	\end{eqnarray}

The equation for \textit{the points where the phases} Sp$_{12}$, Sp$_2$, \textit{and} AF$_2$ \textit{meet}
	\begin{eqnarray}
		&&16A^2x^6 - 16A^2x^4 - 8ALx^3 + 2ALx + L^2 = 0, \nonumber \\
		&&B = -2Ax^4 +\frac{L}{2}x, \nonumber \\
		&&K = -\frac{4Ax^5 - 8Ax^3 - Lx^2 + 4Ax - L}{2x}.
		\label{eq33}
	\end{eqnarray}
If $A \rightarrow -\infty$, then $K  \rightarrow 1$ and $B  \rightarrow L-2A$ ($L > 0$).
At $A = -\frac49 \sqrt6 L$ the first equation has two equal real solutions $x=\frac{\sqrt6}{4}$ ($B = \frac {\sqrt6}{4} L$, $K = \frac{29\sqrt6}{36}$). 
At $-\frac{4}{9}\sqrt6 L < A < 0$	($L > 0 $) the ground state phase diagram is of the type shown in Fig.~\ref{fig2}(a).

\textit{The points where two first order phase transition lines meet} (points of minimal value of $\alpha$ for Sp$_2$ phase) are determined from the following set of equations
	\begin{eqnarray}
		&&4Ax[12x^4 - 8(y^2+y+1)x^2 - 2y^3-y^2+2y+1] \nonumber \\
		&&~~~-L(12x^2-y^2+2y+2) = 0, \nonumber \\
		&&A^2[80x^6 - 48(y^2+y+1)x^4 - (8y^3+4y^2-8y-4)x^2]  \nonumber \\
		&&~~~-16ALx^3 - L^2 = 0,~~~~x = \cos\frac{\alpha_1}{2},~~ y = \cos\alpha_2,	\nonumber \\
		&&B = \frac{L^2x(y+2)}{2y[8Ax^5 - 8Ax^3 - 2Lx^2 - 2A(y^2 - 1)x - L]}, \nonumber \\ 
		&&K = \frac{8Ax^3 - 4Ax + 2Bx + L}{2x}. 
		\label{eq34}
	\end{eqnarray}
At $A =  -\frac49 \sqrt6 L$ ($L > 0$), $x=\frac{\sqrt6}{4}$, $y = -1$ and the region for Sp$_2$ is reduced to a point.

The equation for \textit{lines of fixed $\alpha$ for the phases} Sp$_2$ reads
\begin{equation}
	K = \frac{8AB\cos^3\alpha - 4B^2\cos^2\alpha + L^2}{4B\cos^2\alpha}.
	\label{eq35}
\end{equation}

Equation 
	\begin{equation}
		8A\cos^3\frac{\alpha}{2} + 2(B-K-2A)\cos\frac{\alpha}{2} + L = 0
		\label{eq36}
	\end{equation}
(for the phase Sp$_{12}$) has two real roots $0 < \alpha < \pi$ for $L > 0$ if 
	\begin{eqnarray}
		&&27A + 4(B - K - 2A)^3 < 0, \nonumber \\
		&&A > 0,~~10A + B - K > 0, \nonumber \\
		&&4A + 2B -2K + L > 0
		\label{eq37}
	\end{eqnarray}

and one real root if
	\begin{equation}
		4A + 2B -2K + L < 0.
		\label{eq38}	
	\end{equation}

If the phase with the energy $E = -K - A$ exists, then
	\begin{equation}
		B < 0,~~K > L + B,~~K > -\frac{L^2}{4B} > 0. 
		\label{eq39}
	\end{equation}

If $A < 0$, $ L > 0$ this phase cannot exist in the region where Sp$_{12}$ phase is defined ($4A + 2B -2K + L < 0$).

	\begin{figure}[]
		\begin{center}
			\includegraphics[scale = 0.9]{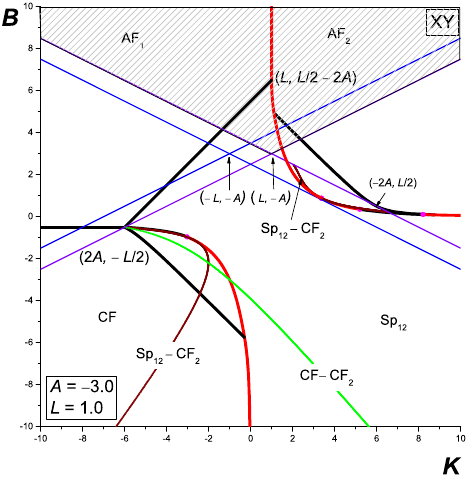}
			\caption{To the proof that the phases CF$_1$ and CF$_2$ cannot exist at $A < 0$. The definition regions for these phases (see Table~\ref{table1}) are shown by blue and violet lines, respectively. The shaded area is the upper part of the definition region for the CF$_2$ structure. The curves where energies of phases CF$_2$ and Sp$_{12}$ (CF$_2$ and CF) are equal are outside the region Sp$_{12}$ (CF).}
			\label{fig12}
		\end{center}		
	\end{figure}

Let us prove that there is no minimum energy region for the phase CF$_1$ (CF$_2$). The energy difference between phases CF$_1$ (CF$_2$) and AF$_2$ is equal to 
    \begin{equation}
        \frac{(K \pm L)^2}{4(A + B)} \pm L - B - (-K - A - 2B) = \frac{(2A + 2B + K \pm L)^2}{4(A + B)}.
        \label{eq40}
    \end{equation}
This value is negative only when $B < -A$. But this is beyond the upper region where the phase CF$_1$ (CF$_2$) is defined (see Fig.~\ref{fig12}). Thus, this phase cannot have the minimum energy at $B > -A$.

At $L > 0$ the energy of the CF$_2$ phase is less than the energy of the phase CF$_1$ (in the region where both phases are defined).

\section*{$\mathbf{A > 0}$, $\mathbf{L > 0}$ case}

\textit{The boundary between spiral} Sp$_{12}$ \textit{and antiferro} AF$_1$ \textit{phases} (a straight line) is determined by the following equations 
	\begin{eqnarray}
	&&4Ax^3 + 8Ax^2 + 4Ax - L = 0, \nonumber \\
	&&B = K - 6Ax^2 - 4Ax;~~ x = \cos\frac{\alpha}{2}.
	\label{eq41}
	\end{eqnarray}

If $A \rightarrow  \infty$ then $B \rightarrow K - L$, $\alpha \rightarrow \pi$.

\bibliography{Collection_2019}

\end{document}